# Analysis of short-term Polarization Stability using Allan Variance

Borja Vidal

Nanophotonics Technology Center, Universitat Politècnica de València, 46022 Valencia, Spain

*Abstract*—The use of Allan variance to characterize the stability of optical signals affected by stochastic polarization fluctuations and the identification of the underlying power law noise processes is explored. Allan variance can ease the comparison in terms of polarization stability of optical systems affected by polarization fluctuations and define a near-optimum integration interval to reveal trends. Experiments with different optical sources show that white noise and random walk terms can be observed. This information can easily be used to systematically define, in real time, the denoising strategy in polarization-based sensing and for the optimization of polarization sensitive optical systems instead of the conventional approach relying on heuristics or information criteria.

## I. Introduction

ELECTROMAGNETIC waves are defined by three physical degrees of freedom, namely, their intensity, frequency and polarization. The performance of light-based systems relies on the precise determination of these properties. Optical amplitude noise processes have been thoroughly studied [1-2]. This knowledge allows proper design of optical networks and photonic systems. Similarly, the frequency stability of crystals and laser oscillators has been extensively researched [3-4], enabling ultra-precise measurements in several fields of physics as well as opening new engineering applications such as satellite navigation, and network synchronization.

The evolution of the state of polarization (SOP) of an optical signal, defined in the Jones or Stokes spaces, has been well studied [5]. In fiber optics, the polarization state of light fluctuates with time due to variations in wavelength or changes in birefringence which is very sensitive to any nonsymmetric perturbation about the fiber axis that can be caused by varying environmental conditions such as temperature or mechanical changes. Considerable attention has been devoted to the study of polarization changes with frequency and, similarly, fiber length [5-9]. However, the time evolution of the SOP is harder to characterize given the lack of homogeneity [10-14].

To quantify the polarization stability of a signal, some common metrics are the rate of change of the SOP, i.e. the angular velocity in the Stokes space [12-14] for a fixed input SOP, which is suitable to quantify the magnitude of polarization transients amid polarization drifts; and the mean squared error of Stokes parameters [11], which enables the assessment of changes from a given SOP.

Here, a tool, originally developed to address issues in frequency stability and synchronization between atomic clocks, is applied to study the character of noise terms affecting polarization fluctuations in optical systems. Allan deviation, or variance, is a time-domain analysis technique designed for the characterization of noise by measuring the heterogeneity of its change across time. It is based on representing the root mean square (RMS) of an error signal as a function of averaging times which has been extensively used to characterize frequency stability of oscillators [3]. This technique allows the characterization of the underlying random processes that drive stochastic fluctuations.

## II. Polarization stability

The SOP of a light signal can be defined in the Stokes space as a vector, $\hat{s}$, which represents a point on the Poincaré sphere [5]. Changes in the SOP at the output of a system can be due to random changes in the birefringence caused by mechanical or thermal fluctuations but also to changes in the wavelength of the optical signal or phase noise.

Thus, fluctuations affecting the measured SOP, $\hat{s}_{meas}$, can be represented as errors on the polarization vector. Their aggregated magnitude, as a measure of the strength of the fluctuations, can be collapsed in a single factor by measuring how far each sample has deviated from its ideal position on the Poincaré sphere, $\hat{s}_{reference}$. This mean squared error, $\hat{s}_{error}$, [11] quantifies the level of fluctuations experienced by the signal. Thus, the amount of stochastic fluctuations on the instantaneous Stokes parameters can be estimated calculating an error vector as the difference between the measured and ideal SOPs, as shown in Figure 1.

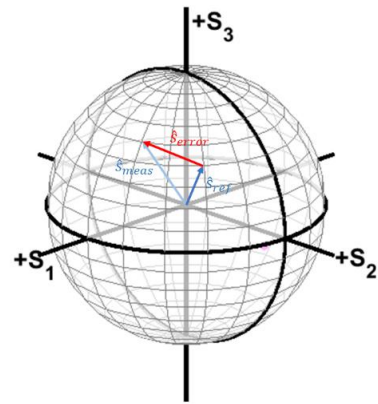

Fig. 1. Concept of polarization error vector (red) on the Poincaré sphere.

This work was supported in part by the Spanish Ministerio de Ciencia e Innovación–Agencia Estatal de Investigación under Project PID2019-111339GBI00.

A factor can be defined to measure the accumulated error as a ratio of the root mean square (rms) value of all the error vectors, averaged over N measurements [11],

$$|\hat{s}_{error}^{rms}| = \sqrt{\left(\frac{1}{3N}\right)\sum_{i=1}^{N}(s_{1,i}-s_{1,ref})^2 + (s_{2,i}-s_{2,ref})^2 + (s_{3,i}-s_{3,ref})^2} \quad (1)$$

where $s_{1..3}$ are the normalized Stokes parameters. This parameter eases comparison among systems in terms of polarization stability as, for example, the conceptually similar error vector magnitude (EVM) parameter provides a system level metric which is used to quantify the performance of digital modulations under varied impairments.

### III. 3. ANALYSIS OF POLARIZATION STABILITY USING ALLAN DEVIATION

Some stochastic processes with scale-free dynamics have power spectral densities that follow a power law, i.e. their spectral density can be approximated by a sum of terms each varying as an integer power of frequency. A power law process has a spectral density of the form,

$$S(f) = \sum_{\alpha=-2}^{2} h_\alpha f^\alpha \quad (2)$$

where some terms of the summation are usually dominant. Thus, the PSD of each power-law noise process can be specified by its slope on a log-log plot for a given range of frequencies and its amplitude.

Through the estimation of the polarization stability provided by $|\hat{s}_{error}^{rms}|$, Allan deviation can be used to investigate the power-law noises affecting polarization fluctuations. Let be a set of $N$ consecutive data points showing the instantaneous SOP error, $|\hat{s}_{error}^{rms}(t)|$, each sampled at rate $T_s$. We can group $n$ samples (with $n < N/2$) in a cluster. The temporal duration of each cluster, or correlation time, is $\tau = n \cdot T_s$. The average for a cluster which starts from the $k$ data point and contains $n$ subsequent points is

$$\bar{y}_k[n] = \frac{1}{\tau}\sum_{i=k}^{k+n}|\hat{s}_{error}^{rms}(i)| \quad (3)$$

Thus, Allan variance, $\sigma^2(\tau)$, is defined as half the averaged squared mean of two adjacent clusters,

$$\sigma^2(\tau) = \frac{1}{2}\langle(\bar{y}_{n+1}-\bar{y}_n)^2\rangle \quad (4)$$

where $\langle\cdot\rangle$ denotes the average. It is a positive value and it can only be calculated for $\tau < \frac{N\cdot T_s}{2}$ or equally $n < \frac{N}{2}$. For large values of $\tau$, the number of clusters is low and consequently the statistical error increases. To reduce the large statistical error of standard Allan variance, more clusters can be obtained by overlapping samples among clusters [15]. First, the difference between two clusters each of $n$ samples is calculated. Then each cluster is shifted one sample and the difference calculated again. This process is repeated to obtain a larger set of averages. Thus, the overlapping Allan variance shows better statistical error at the cost that clusters are no longer statistically independent.

Allan variance is related to the power spectral density, PSD, of the random process [16]. This relation is:

$$\sigma^2(\tau) = 4\int_0^\infty S(f)\cdot\frac{\sin^4(\pi f\tau)}{(\pi f\tau)^2}df \quad (5)$$

where $S(f)$ is the PSD of $|\hat{s}_{error}^{rms}(t)|$. From (5) it can be seen that Allan variance is proportional to the total power output of the random process after being filtered by a frequency response given by $\sin^4(\pi f\tau)/(\pi f\tau)^2$. In other words, Allan variance can be calculated in the time domain as a convolution (3) or in the frequency domain as a filter (5). The bandwidth of the filter depends on $\tau$, thus, the different types of underlying noise terms in the signal can be identified and quantified by varying $\tau$. This is usually done through a log-log plot of $\sigma(\tau)$ versus correlation time which allows to sort out noise components by the slopes of Allan deviation [17].

As an example, Fig. 2 shows logarithmic plots of Allan deviation for the change in polarization obtained from three light sources connected to an Optellios polarimeter (PS2300B) through a polarization controller (Fig. 2a), a polarization-maintaining fiber patchcord (50 cm) (Fig. 2b) or a polarization controller plus a differential group delay (DGD) (Fig. 2c). Figure 2 also shows the temporal evolution of Stokes parameters and their Poincaré sphere representation, Fig. 2d-f. Measurements were carried out with a sampling rate of the polarimeter of 200 Hz or 20 Hz (system c) and the number of samples was 10000. The first data point is selected as the reference SOP.

Several piecewise linear regions with different slopes can be observed in the three cases. These straight-lines on the log plot are potential signatures of a power law. From the figure the main types of dominant error can be identified.

The downward slope, with a slope of $-1/2$, can be associated to white noise following Allan analysis theory. Fluctuations on the Stokes parameters of the output signal stem from random changes in the local birefringence of the fiber induced by several coexisting mechanisms such as geometric, twist and stress irregularities under wavelength and environmental changes. Each can be seen as a Gaussian independent random variable with zero mean [6,9]. Coherent interference between the polarization modes suffering the accumulated effect of all these perturbations give rise to high frequency random changes on the SOP with a correlation time much shorter than the sampling time and they can be characterized by a white noise spectrum.

$$S(f) = N_0^2 \quad (6)$$

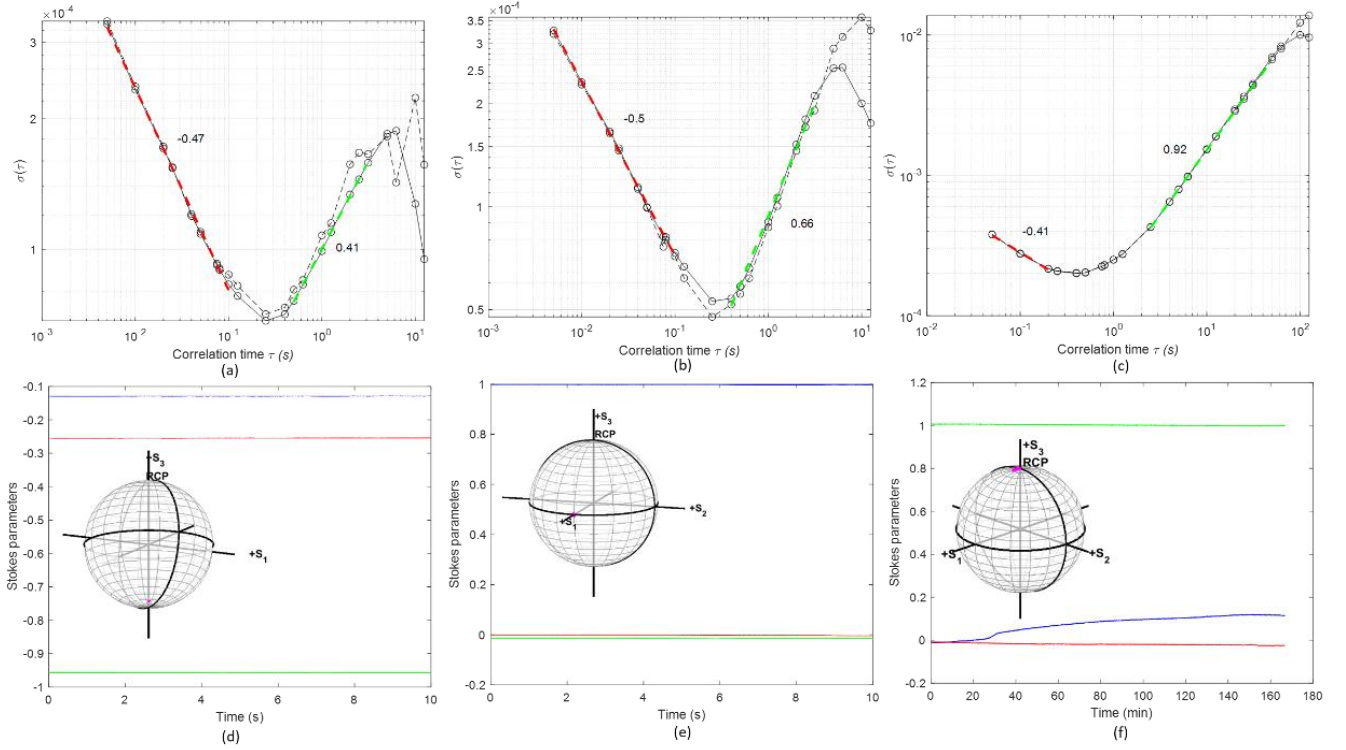

Fig. 2. Log-log plot of Allan deviation of SOP for three optical systems made of an optical source connected to the polarimeter through a polarization controller; dashed: non-overlapping estimator, solid: overlapping estimator; (a) Yenista laser (model 1549.32) (b) Keysight laser (model 81940A); (c) Keysight laser (model 8168F) connected to a DGD module (90 ps). Red and green dashed lines are linear regression fits. The evolution of the sampled Stokes parameters evolution of SOP of each measurement on the Poincaré sphere is shown in (d), (e) and (f).

Substituting (6) in (5), the Allan deviation yields,

$$\sigma(\tau) = \frac{N_0}{\sqrt{\tau}} \quad (7)$$

Thus, the Allan deviation of the white noise term can be derived, which will show a slope of $-1/2$. It is what is observed in the experimental data shown in Fig. 2, whose slope was calculated through a linear regression fit. The quantification of the PSD can be obtained by reading the value at $\tau = 1$ of the linear regression of the data. As an example, the sampled white noise PSD terms for the systems (a) and (b) in Fig. 2 are $N_0^{(a)} = (2.2 \pm 0.5) \cdot 10^{-5}$ and $N_0^{(b)} = (2.3 \pm 0.5) \cdot 10^{-5}$, respectively. These values of the sampled white noise term are associated to the continuous noise which has been filtered by the analog bandwidth of the polarimeter. Additionally, these values should be interpreted as upper bounds since quantization of the Stokes parameters by the polarimeter might introduce an additional white noise term (Annex I).

Figure 3 shows the experimental data for the white noise term of Fig. 2a, simulations of a discrete white noise process with the same standard deviation as the one derived from Allan analysis and a linear fit to the data which has a $R^2$ value of 0.9987.

A second upward slope can be observed in the plots of Fig. 2. It can be associated to a low frequency drift, i.e. noise which changes over longer time frames and therefore starts to affect larger data clusters. Given the range of slopes observed, it can be interpreted as a fractional Brownian motion (fBm) noise [18], i.e. a Gaussian zero-mean nonstationary stochastic process with stationary increments and which is defined by the Hurst exponent, $H$, which goes from 0 to 1. This noise is an extension of ordinary Brownian noise, which is obtained for $H = 1/2$. Values $H < 1/2$ result in a mean reverting process (antipersistent), whereas $H > 1/2$ reflect a process than has a trend (bias). The closest that $H$ is to 1 the greater the degree of persistence or long-range dependence, i.e. the evolution of the SOP shows statistically significant correlations across large time scales. In the study of polarization evolution, it could be associated to a global birefringence. The Hurst exponent can directly be inferred from the slope of Allan deviation on the log plot.

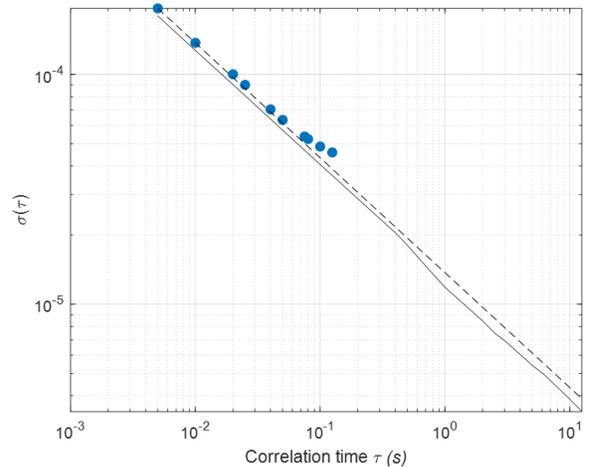

Fig. 3. Overlapping Allan deviation, σ(τ), for the white noise term on a log-log plot: Allan deviation calculated from white noise simulations with parameters from Fig. 2a (solid black); Theoretical linear fit with -0.5 slope (dashed black); Experimental results from Fig. 2a (blue dots)..

In particular, for the system described in Fig. 2a, the slope is around $+1/2$ on the log—log plot following the dynamics of Brownian noise, i.e. the short time future evolution of SOP cannot be predicted since values are uncorrelated. Several models of the polarization evolution in optical fibers with propagation length or frequency rely on having the PMD vector to change orientation randomly as a random walk [5,6]. This random polarization drift can be interpreted as a continuous Wiener process that is being discretized by the polarimeter becoming a random walk. It is a nonstationary process that manifests itself as impulsive changes on the SOP. The random walk (Brownian) nature of the noise means that there is no memory or correlation from one impulse to the next. These discrete random disturbances can be described as a set of random steps on the mathematical space of the Poincaré sphere. The PSD of a random walk noise is [16]

$$S(f) = \frac{K^2}{(2\pi f)^2} \quad (8)$$

The standard deviation associated to a random walk noise can be obtained by substituting (8) into (5). It can be expressed as

$$\sigma(\tau) = K\sqrt{\frac{\tau}{3}} \quad (9)$$

where $K$ is the random walk coefficient. The magnitude of this term can be read from the fit slope line at $\tau = 3$. The value associated to the measurement of Fig. 2a is $K^{(a)} = (1.7 \pm 0.7) \cdot 10^{-4}$. Figure 4 shows a comparison between the measurement of Fig. 2a, the theoretical slope $(+1/2)$ and simulations of Brownian noise. The goodness-of-fit measure through $R^2$ shows a value of 0.9942 pointing out, as in the white noise term, to the plausibility of a power law as a fit of the data.

On the other hand, the measurements shown in Fig. 2b-c show that the Hurst exponents, 0.66 and 0.92 respectively, point out to the presence of a trend, i.e. the system is evolving through a systematic (deterministic) bias rather than being affected by noise, especially in Fig. 2c. This can be attributed to the precession of the SOP around the PMD vector axis under frequency drift at a rate that is governed by the strength of the birefringence. This evolution can be seen in the inset of Fig. 2f as an arc where the system has a strong birefringence (90 ps). Thus, Allan analysis helps to identify and quantify the nature of polarization fluctuations. Thus, the deterministic error component can be searched and corrected to enhance the stability of the system.

In addition, in the systems studied in Figure 2, Allan deviation plots show a sweet spot where standard deviation is minimized due to the averaging of fast-oscillating noise whereas the data are not being corrupted yet by slower polarization drifts. When a moving average analysis is carried out, the window length is critical in the performance of the estimator. Instead of evaluating multiple window lengths to select the near optimal using heuristics or even information criteria such as Akaike Information Criterion (AIC), Allan deviation provides a systematic data-driven method to determine the timescale over which the SOP remains relevant while minimizing white noise corruption of the nonstationary term. This is done without any prior knowledge about the system or its noise model. As an example, for the dataset shown in Fig. 2a, a correlation time of $\tau = 0.25$ allows SOP fluctuations to be minimized.

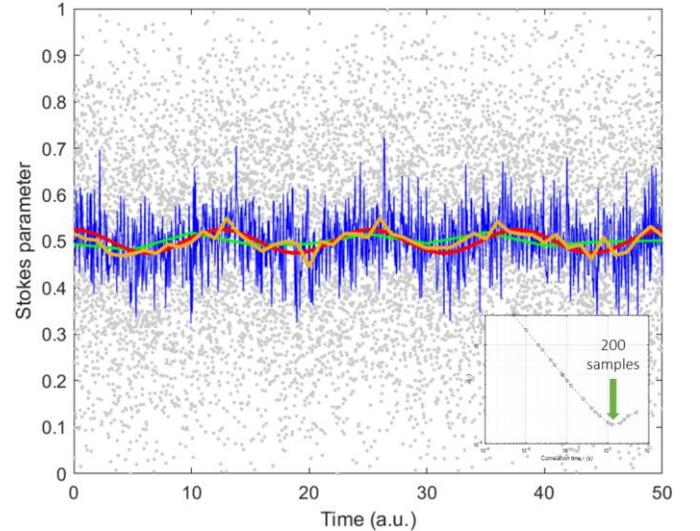

Fig. 5. Simulation of the evolution of a Stokes parameter driven by white Gaussian noise plus a small-amplitude cosine signal (red solid) sampled at 200 Hz. Evolution of the Stokes parameter (grey dots); moving average with window length 10 samples (blue solid); moving average with window length 200 samples, optimum point according to Allan deviation plot, (orange solid); moving average with window length 1000 samples (green solid). Inset: Allan deviation plot of the Stokes parameter simulations.

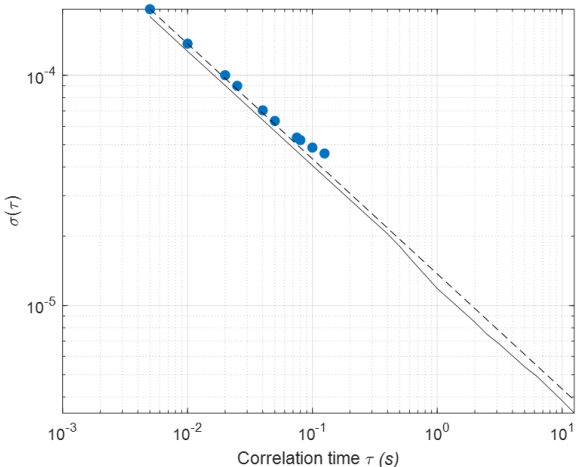

Fig. 4. Overlapping Allan deviation, $\sigma(\tau)$, for the random walk term on a log-log plot: Allan deviation calculated from simulations of Brownian noise on a sphere with the same coefficient than measurements from Fig. 2a (solid black); Theoretical linear fit with 0.5 slope (dashed black); Experimental results shown in Fig. 2a (blue dots).

Figure 5 shows simulations of a small amplitude cosine signal embedded in additive white Gaussian noise. Allan analysis shows that the near-optimum length of the moving average filter is 200 samples. This value is used to define the window length of the optimum moving average filter that better retrieves information buried in noise for the dataset, i.e. a better estimation of the underlaying deterministic component. As a comparison, the sum of squared estimate of errors (SSE) for the measurement (50000 samples) goes from 2000 without averaging to 9 for a moving average of 200 samples. For an averaging window of 10 and 1000 samples, the SSE is 174 and 19, respectively.

Figure 6 shows simulations and measurements of a Stokes parameter under noise and polarimeter quantization and how the underlying noise term (random walk) is revealed when quantized white noise is removed through optimum moving average filtering with the window length given by Allan analysis. Since white noise is larger than quantization noise, digitalization introduces an offset on the Allan curve and consequently the noise coefficients associated to each noise term but it does not significantly change the optimum moving average filter length.

Finally, the comparison of the analysis of each Stokes parameter provides information about the isotropic nature of the noise terms on the Poincaré sphere. Fig. 7a shows Allan deviation for each of the Stokes parameters for the same datasets presented in Fig. 2a. For these systems, the white noise term is rather similar between the Stokes parameters, however, it can be seen that it not exactly the case for the fractional Brownian noise. This suggests that, in systems with small total fiber length, the fBm component can be anisotropic, i.e. not all possible orientations of the error term of the Stokes vector are equally likely. On the contrary, in long fibers, it was shown that the SOP evolution can be seen as random walk equally likely in all directions [12].

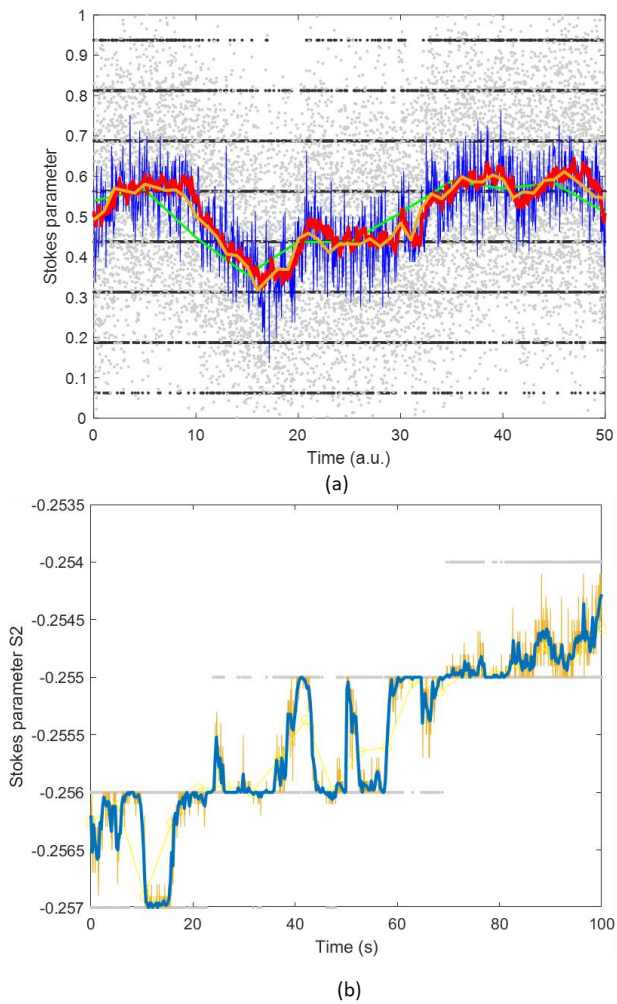

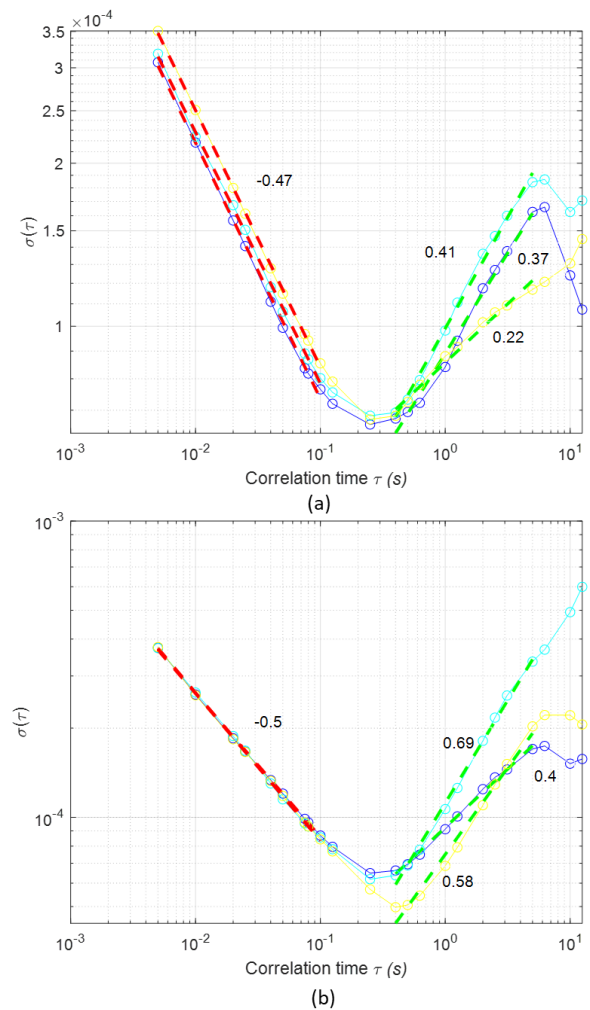

Fig. 6. (a) Simulation of a random walk evolution of a Stokes parameter (red), measurement in a polarimeter under white noise (light grey) plus quantization (dark grey dots), estimations of the evolution with different moving average windows lengths (blue: 10 samples; green: 1000 samples; orange: optimum value, 200, given by Allan analysis); (b) Experimental evolution of the quantized S2 Stokes parameter (grey dots); estimation of the unknown evolution through the moving average with optimum window length (50 samples or 0.25 s) (blue solid); estimation through a moving average with window length 10 samples (orange solid)..

Fig. 7. Overlapping Allan deviation for each Stokes parameters: S1 (blue), S2 (cyan), S3 (yellow). (a) System described in Fig. 2a; (b) System described in Fig. 2b.

## IV. Conclusion and discussion

The use of a simple data-driven method available in most scientific software packages, Allan deviation, for the analysis of stochastic polarization time fluctuations as a tool to study underlaying noise models and the optimization of denoising in polarization measurements has been proposed and studied. Allan analysis has shown that short-time SOP evolution scale with multiple scaling rules rather than following a global scaling rule. Observations are consistent with the hypothesis that the SOP evolves following a power law with different regimes and different regularity patterns can be distinguished on the data: a white noise term, i.e. a mean reverting (anti persistent) process; a random walk; and a fractional Brownian process that in the extreme develops in a trending (persistent) process. These short-time temporal noise terms agree with previously reported statistical models for the evolution of polarization with fiber length and optical frequency which suggest that the changes in the PMD vector are driven by changes in orientation following a Brownian motion and a white rotation process [5-6].

Allan deviation provides a systematic approach to the selection of the time window length for averaging over which measurements remain relevant. These noises, following a power law, are scale invariant, or, in other words, they are self-similar. This implies that there are no restrictions on the selection of the sampling frequency since the characteristics of the processes are conserved at different time scales.

An alternative path for analyzing the regularity of polarization fluctuations could be in the framework of multifractal analysis (or scaling analysis). SOP fluctuations can be seen as a multifractal behavior with several monofractal colored noise terms dominant. Fractal structures similarly allows the identification of power law processes. However, multifractal tools such as leader wavelets and multifractal detrended fluctuation analysis do not provide a simple derivation of the optimum time window for denoising and they are not as easy to use.

Allan analysis can contribute to advance the study of systems where the time evolution of the SOP is relevant such as in the optimization of the averaging in polarization measurements, comparison of optical system in terms of polarization stability, modelling and quantifying noise stemming from various sources as well as to provide new insights on the polarization behavior of optical systems. Different applications may benefit from Allan analysis such as the optimization of polarimetric sensing, especially at low signal power, [19], as in remote sensing [20] and astronomy [21]; the calibration of polarization elements, characterization of materials [22] and characterization of polarization speckle [22]; as well as the alignment of polarization-critical optical systems, as in quantum key distribution systems [24], microwave photonics [25] and fiber-optic gyroscopes [26]. This tool might strengthen the accuracy of these applications with an analysis that can be done on-the-fly to determine optimal parameters.

## ANNEX I

### A. Effect of quantization noise on Allan analysis

The rounding errors of the Stokes parameters introduced by the polarimeter as it digitally encodes analog inputs will introduce a quantization. This nonlinear process, as in any analog-to-digital conversion, can be modelled as a quantization noise. i.e. an additive white noise. For polarimetric measurements, the variance of quantization noise is given by

$$\sigma_q = \frac{1}{\sqrt{12}} \cdot \frac{1}{2^{B-1}} \quad (A.1)$$

where B is the number of bits of the uniform quantizer.

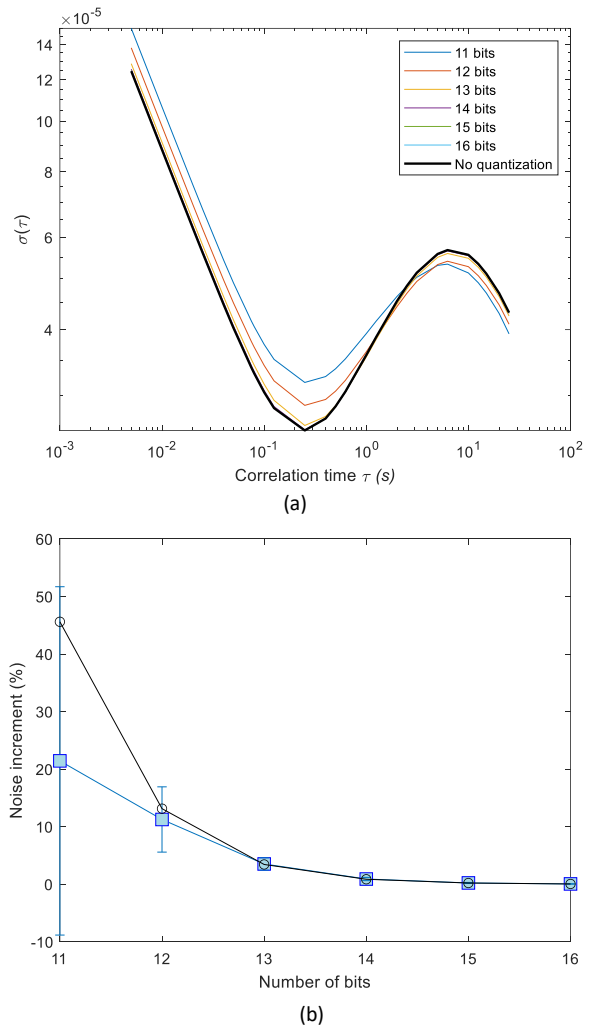

Fig. A.1. (a) Allan deviation analysis of simulated samples for different number of uniform quantization levels; b) Relative increment of the white noise term as a function of the number of quantization levels. Blue: simulations; black: theoretical increment from quantization noise.

The effect of quantization on Allan deviation can be shown in Fig. A.1 where simulations of white plus Brownian noise terms have been taken into account. The impact of quantization noise on the total white noise depends on the strength of the amplitude of the white term in relation to the least significant bit of the quantization, i.e. the quantization step. The smaller the analog white noise variance the higher the impact of quantization noise.

As it can be seen, the effect of quantization noise on the Allan plot departs from conventional Allan analysis of oscillators or gyroscopes where quantization noise on the frequency of a source or the gyro output manifests as a characteristic noise term with a -1 slope. Here, quantization of the Stokes parameters results in an amplitude-like noise contribution.